# Decomposition of solid hydrogen bromide at high pressure

Defang Duan, Fubo Tian, Xiaoli Huang, Da Li, Hongyu Yu, Yunxian Liu, Yanbin Ma, Bingbing Liu, Tian Cui*

State Key Laboratory of Superhard Materials, College of Physics, Jilin University, Changchun 130012, People's Republic of China

*Electronic address: cuitian@jlu.edu.cn

## Abstract

The stability of different stoichiometric $H_nBr$ (n=1-7) compounds under pressure are extensively studied using density functional theory calculations. Five new energetically stable stoichiometries of $H_2Br$, $H_3Br$, $H_4Br$, $H_5Br$, and $H_7Br$ were uncovered at high pressure. The results show that HBr is stable below 64 GPa, then decomposes into new compound $H_2Br$ and $Br_2$ molecular crystal. For $H_2Br$ and $H_3Br$ compounds, they were found to become stable above 30 GPa and 8 GPa, respectively. In addition, we accidentally discovered the triangular $H_3^+$ species in $H_5Br$ compounds at 100 GPa. Further electron-phonon coupling calculations predicted that hydrogen-rich $H_2Br$ and $H_4Br$ compounds are superconductors with critical temperature of superconductivity $T_c$ of 12.1 K and 2.4 K at 240 GPa, respectively.

## Introduction

The discovery of superconductivity with an extraordinarily high $T_c$=190 K in $H_2S$ at 200 GPa has generated considerable excitement in the scientific community.[1] It was theoretically predicted by our group,[2] prior to the observations of this high $T_c$, that $(H_2S)_2H_2$ is thermodynamically stable up to 300 GPa, and becomes metallic with $H_3S$ molecular unit above 111 GPa and superconducting with a $T_c$ in the range of 191 K and 204 K at a pressure of 200 GPa. The stability of the $H_nS$ series of compounds at high pressures was then substantially explored, and concluded that $H_2S$ really decomposes into S and new compound $H_3S$ above 50 GPa, and $H_3S$ is stable at least up to 300 GPa.[3-5] They believed that the $H_2S$ sample exhibiting superconductivity at 190 K comes from the *Im-3m* phase of $H_3S$. Later, other group IVB hydrides (H-Se, H-Te and H-Po) were predicted to be high-temperature superconductors at high pressure.[6-9] Besides these efforts, the stability and superconductivity of other hydrides at high pressure have also attracted great attention.

Solid hydrogen bromide (HBr) shows three crystalline phases at low temperature and ambient pressure.[10, 11] The lowest temperature phase (phase III) has an orthorhombic ordered structure with *Cmc*$2_1$[10]. With increasing temperature, phase III transforms to phase II (*Cmca*) at 90 K, then to phase I (*Fm-3m*) at 114 K, and phases I and II are orientationally disordered[11]. Phase transitions at high pressure have been investigated experimentally[12-14]. At room temperature, phase I transformed to phase III at 13 GPa, then to phase IV with symmetrized hydrogen bonds at 39 GPa. Furthermore, HBr was considered to decompose into $Br_2$ and possibly $H_2$ after hydrogen-bond symmetrization.[13] In that experimental research, the Raman spectra of $Br_2$ molecules were observed, but Raman signals of $H_2$ molecules were not detected due to much weak Raman scattering intensity compared to $Br_2$ molecules.

Several theoretical studies have been performed to understand the hydrogen-bond symmetrization and structural stability of solid HBr under pressure.[15, 16] *Ab initio* molecular dynamics simulation revealed a spontaneous formation of $H_2$ molecules with monatomic Br lattice above 40 GPa.[16] In addition, they also showed that the disorder at higher pressure leads to cooperative proton-transfer dynamics. The mechanism of hydrogen-bond symmetrization were well understood and higher pressure structures were predicted by *ab initio* calculation.[17, 18] Moreover, they suggested that HBr decompose into $H_2$ molecules and monatomic Br above 120 GPa

or 196 GPa.

The previous theoretical studies only discuss the enthalpies of $H_2S$ relative to $H_2$ and $Br_2$ at high pressure[17, 18]. However, the stability of HBr against decomposition into other stoichiometric compounds, e.g., $H_2Br$, $H_3Br$ or $H_5Br$, has not been investigated up to now. Therefore, in the present study, we aim to elucidate the high-pressure stability of different stoichiometric $H_nBr$ (n=1-7) to determine the pressure and products of HBr decomposition. Results show that HBr decompose into new compound $H_2Br$ and $Br_2$ molecular crystal above 64 GPa. In addition, we accidentally discovered the triangular $H_3^+$ species in $H_5Br$ compounds above 100 GPa. Moreover, the $T_c$ for hydrogen-rich $H_2Br$ and $H_4Br$ compounds at 240 GPa is 12.1 K and 2.4 K, respectively.

**Computational Methods**

To obtain stable structures for $H_nBr$ (n=1-7), the CALYPSO structure prediction method are used based on a particle swarm optimization algorithm combined with *ab initio* total-energy calculations as implemented in the CALYPSO code.[19, 20] The effectiveness of this method has been demonstrated by the successful applications in predicting high-pressure structures of various systems.[21, 22] The underlying structure relaxations are performed using the projector augmented waves method[23], as implemented in the Vienna *ab initio* simulation package VASP code[24]. The generalized gradient approximation (GGA) of Perdew-Burke-Ernzerhof (PBE)[25] is adopted to describe the exchange-correlation potential. The energy cutoff 1000 eV and k-mesh of $2\pi \times 0.03$ Å$^{-1}$ within the Monkhorst-Pack scheme are chosen to ensure that the total energy are well converged to better than 1 meV/atom. The Bader analysis and ELF are also calculated using VASP code.

Lattice dynamics and superconducting properties are calculated using density functional perturbation theory as implemented in the QUANTUM-ESPRESSO code [26]. The norm-conserving potentials[27] are used and convergence tests give a suitable value of 80 Ry kinetic energy cutoff. The q-point mesh in the first BZ of $5 \times 5 \times 5$ for $H_2Br$-*Cmcm* and $3 \times 3 \times 5$ for $H_4Br$-*P6$_3$/mmc* are used in the interpolation of the force constants for the phonon dispersion curve calculations. A denser k-point mesh 20 × 20 × 20 for $H_2Br$-*Cmcm* and 12 × 12 × 20 for $H_4Br$-*P6$_3$/mmc* are adopted to ensure k-point sampling convergence with a Gaussians width of 0.03 Ry.

## Results and Discussion

The crystal structure predictions are performed by considering simulation sizes ranging from 1 to 4, 6, and 8 formula units per cell (fu/cell) for $H_nBr$ (n=1-3) and from 1 to 4 fu/cell for $H_nBr$ (n=4-7) at 20, 50, 100, 200 and 300 GPa. The stability of $H_nBr$ compounds is quantified by constructing the thermodynamic convex hull at the given pressures, which is defined as the enthalpy of formation per atom for the most stable structures of $H_nBr$ at each stoichiometry. The enthalpy of formation per atom at each pressure is calculated using the following formula: $h_f$ ($H_nBr$)=[$h(H_nBr)−h(Br_2)/2−nh(H_2)/2$]/(n+1). Any structure with the enthalpy of formation lies on the convex hull is considered to be thermodynamically stable and synthesizable experimentally[28]. The convex hulls of $H_nBr$ at 40, 60, 70, 100, 200, and 300 GPa are depicted in Fig. 1.

At 40 GPa (Fig. 1a), the enthalpy of formation for HBr, $H_2Br$, $H_3Br$ and $H_7Br$ fall on the convex hull, which indicates that these compounds are thermodynamically stable at this pressure. In addition, HBr has the most negative enthalpy of formation, which is consistent with the fact that HBr exists at a low pressure range. As pressure increased, $H_2Br$ became the most stable stoichiometry, and HBr began to deviate from the tie-line at 70 GPa (Fig. 1c). This phenomenon clearly suggests that HBr is instable and will decompose into $H_2Br$ and Br2 molecules via the reaction 2HBr→$H_2Br$+Br. Furthermore, the decomposition enthalpies of HBr relative to that of the $H_2Br+Br_2$, $H_3Br+Br_2$, $H_5Br+Br_2$, $H_5Br+Br_2$, and $H_2+Br_2$ as a function of pressure have plotted in Fig. S1 in the Supplemental Material to confirm the pressure of decomposition. It is more clearly seen that HBr decomposes into $H_2Br$ and $Br_2$ molecules above 64 GPa. This phenomenon is consistent with experimental observations of HBr molecular dissociation into $Br_2$ molecules at 43 GPa and room temperature. In that experiment, they observed the Raman spectra of $Br_2$ molecules, but did not detect the Raman signals of $H_2$ molecules. The differences between calculated and measured pressures of decomposition could be attributed to many factors, e.g., absence of temperature effects in the calculations and limitations in pseudopotential-DFT approaches.

The pressure-composition phase diagram of $H_nBr$ (n=1-7) is depicted in Fig. 1(g) and shows the stability ranges of the stable stoichiometries ($H_2Br$, $H_3Br$, $H_4Br$, $H_5Br$, and $H_7Br$). It is shown that HBr is stable from 0 GPa to 64 GPa, then it becomes unstable and decomposes into $H_2Br$ and $Br_2$ molecules. $H_2Br$ becomes stable in the pressure range from 30 GPa to 140 GPa, and from 240 GPa to 300 GPa. $H_3Br$ and $H_5Br$ become stable at pressures of 8 GPa and 70 GPa, remain stable up to 180 GPa and 280 GPa, respectively. $H_4Br$ and $H_7Br$ is only stable in a pressure range from 30 to 60 GPa, and from 240 to 300 GPa. The calculations show that $H_6Br$ is unstable at every pressure considered. The structures of $H_nBr$ compounds are all found to be dynamically stable within their stable pressure ranges expect for $H_5Br$-$Pmn2_1$ phase which will be discussed in detail below. Selected phonon band structure and phonon density of states (DOS) are provided in the Supplementary Material.

The selected structures for stable stoichiometries of $H_nBr$ are shown in Fig. 2. HBr is stable in $Cmc2_1$ (0 GPa–25 GPa) and $Cmcm$ (25 GPa–64 GPa) structures up to 64 GPa, which is in agreement with the previously theoretical results[17, 18]. $H_2Br$ is found to be stable in $C2/c$ structure (see Fig. S2 in the Supplementary Material) in the pressure range from 30 to 180 GPa. This structure consists of two HBr zigzag sheets and one $H_2$ molecules straight chain. The HBr sublattice is similar to the pure HBr phase IV,[17] with H atoms occupying midpoint of two neighboring Br (hydrogen bond symmetrization). Then, it is unstable up 240 GPa, a high symmetry structure with $Cmcm$ space group occurs. In this structure, bromine atoms form tubular network that trap $H_2$ molecules units (H-H distance of 0.813 Å at 240 GPa) in channels along the z axis, as shown in Fig. 2a. This arrangement is comparable to the guest-host clathrate structures found in methane hydrate (MH-III)[29].

A $P2_12_12_1$ structure is favored for $H_3Br$ in the pressure range from 8 to 100 GPa, This structure consists of H-bonded HBr zigzag sheets and $H_2$ molecules sheets forming sandwiched configuration (see supplementary Fig. S3). At 10 GPa, the covalent bond length of Br–H is 1.509 Å and hydrogen bond length of H⋯Br is 1.998 Å. When the pressure up to 30 GPa, hydrogen bond symmetrization happens, bond lengths of H⋯Br and H–Br are approximate equal with 1.64 Å, and H–Br–H angle 97.8 °. Further compression to 100 GPa, $P$-1 structure occurs consists of H–Br zigzag chains with H–Br–H angle 89.2°, as shown in Fig. 2b. The predicted $H_4Br$ prefers a hexagonal $P6_3/mmc$ structure from 240 to 300 GPa, as depicted in Fig. 2c. In this structure, H-Br forms 3D network that trap $H_2$ molecules arranged in a straight chain. There are three different $H_2$ pairs with distance of 0.73 Å, 0.76 Å and 0.77 Å. On the other hand, $H_7Br$ was found to be stable in a structure with $P2_1/m$ symmetry, consisting of symmetrization H–Br zigzag chains and two $H_2$ molecules sheets, as depicted in Fig. 2f.

$H_5Br$ is stable with $C2/c$ consisting of symmetric hydrogen bond zigzag chains and $H_2$ molecules sheets above 70 GPa (see supplementary Fig. S2b). Then $Pmn2_1$ structure becomes energetically favorable between 100 GPa and 200 GPa. This structure consists of a $[H_3][Br][H_2]$ complex, as shown in Fig. 2e. Each Br atom contacts with three H atoms in $H_3$ unit and $H_2$ units fill in grid interval. However, this structure is dynamically instable up to 140 GPa (see supplementary Fig. S5). A monoclinic $Cc$ structure which also contains $H_3$ unit was found at about 100 GPa (Fig. 2d). The enthalpies of this structure is lower than that of $C2/c$ above 100 GPa and slight higher than that of $Pmn2_1$ (see supplementary Fig. S6). Therefore, the phase sequence with increasing pressure for $H_5Br$ is follows: from $C2/c$ to $Cc$ at 103 GPa, then to $Pmn2_1$ at 140 GPa, which is stable at least up to 280 GPa. For $Cc$ phase at 120 GPa, the bond lengths of H–H within $H_3$ unit are 0.850 Å, 0.925Å, 0.975 Å and bond angles are 53.1°, 60.5°, 66.4°. For the case of $Pmn2_1$ phase at 140 GPa, $H_3$ unit forms an approximate equilateral triangle with H–H length of 0.903 Å, 0.903 Å, 0.911 Å and bond angles 59.7, 59.7, and 60.6°. Analysis of ELF values of $Cc$ and $Pmn2_1$ show that there exists covalent bond in $H_3$ and $H_2$ molecules, and no-covalent bonding between H and Br atoms (Fig. 3). On the basis of Bader theory, the charges for $H_3$, $H_2$ and Br unit are +0.397, -0.055 and -0.342 for $Cc$ and +0.397, -0.061 and -0.336 for $Pmn2_1$, respectively. It is seen that the charge transfer from H atoms to Br atoms illustrate the ionic nature with the notation of $[H_3]^+[Br]^-[H_2]$.

Further analysis of the ELF for $Cc$ phase as function of pressure, we clearly see the forming process of $H_3^+$ species, as shown in Fig. 3a-c. At 60 GPa, the highest ELF vales are found in the $H_2$ molecules (Fig. 3a). In addition, the ELF values in the region of the H1-Br is above 0.8 suggesting covalent bond and in the region of H1–H2 and H1–H3 are close to 0.5 indicating no bonding. With increasing pressure, the ELF values along the H1–Br bond decreases, push the H1 atoms close to the H2 and H3 atoms resulting in the increase of the ELF values along the H1–H2 and H1–H3 direction (Fig. 3b). At 140 GPa, the H1 and Br ELF basins become virtually disconnected and the values in the region of H1, H2 and H3 is above 0.8 forming $H_3$ molecule unit (Fig. 3f). This corroborates the notion that the application of high pressure leads to an increase of the ionicity of H–Br bond and subsequent effective self-dissociation into $Br^-$ anions and isolated $[H_3]^+$ cations.

The calculated electronic band structure and projected density of states (DOS) of $H_2Br$-*Cmcm* and $H_4Br$-*P6₃/mmc* at 240 GPa are demonstrated in Fig. 4. Note that the calculated valence bandwidths are very broad and show strong hybridization between the Br and H orbitals. The band structures reveal metallic character with band gap closure. For $H_2Br$-*Cmcm*, analysis of the ELF shows a high value ~ 0.9 between two H atoms within the unit, indicating a strong covalent bonding feature (Fig. S7a). In addition, the ELF between Br and Br atoms is about 0.5, suggesting a atomic feature. For the case of $H_2Br$-*Cmcm*, a high ELF value of 0.9 between two H atoms within pairs indicates a strong covalent bonding feature and the ELF value of 0.75 between H and Br suggests weak covalent bonding feature (Fig. S7b). The phonon band structure and phonon density of states (PHDOS) for *Cmcm*-$H_2Br$ and *P6₃/mmc*-$H_4Br$ at 240 GPa are shown in Fig. 5. The absence of imaginary frequency modes indicates that both structures are stable. The heavy Br atoms dominate the low frequencies, the H–Br wagging vibrations contribute to the intermediate frequency region and the high frequencies are mainly due to vibrations of H-H stretching in $H_2$ molecular units.

To explore the superconductivity, we calculate the electron-phonon coupling (EPC) parameter λ, the logarithmic average phonon frequency $\omega_{log}$, and the Eliashberg phonon spectral function $\alpha^2F(\omega)$ of $H_2Br$-*Cmcm* and $H_4Br$-*P6₃/mmc* at 240 GPa. The resulting λ of $H_2Br$-*Cmcm* and $H_4Br$-*P6₃/mmc* is 0.51 and 0.35 at 240 GPa, respectively, indicating that the EPC is relatively weak. The $\omega_{log}$ calculated directly from the phonon spectrum is 919 K and 1302 K for $H_2Br$-*Cmcm* and $H_4Br$-*P6₃/mmc*, respectively. The spectral function $\alpha^2F(\omega)$ and the integrated λ as a function of frequency are predicted in Fig. 5. It is found that for $H_2Br$-*Cmcm*, the contribution from the low frequency Br translational vibrations constitutes 45.1% of the total λ, intermediate frequency H-Br wagging vibrational modes make up a section of 43.1% and the remaining 11.8% is derived from the high frequency stretching vibrations from $H_2$ unit. For *P6₃/mmc*-$H_4Br$, the low-frequency translational vibrations, the intermediate-frequency wagging modes and high-frequency stretching modes contribute 22.9%, 71.4% and 5.7% to the EPC, respectively. The $T_c$ is estimated from the Allen-Dynes modified McMillan equation[30] $T_c = \frac{\omega_{log}}{1.2}\exp[-\frac{1.04(1+\lambda)}{\lambda-\mu^*(1+0.62\lambda)}]$, where $\mu^*$ is the Coulomb pseudopotential. Using the $\mu^*$=0.1~0.13, the critical temperature $T_c$ for the H$_2$Br-$Cmcm$ and H$_4$Br-$P6_3/mmc$ at 240 GPa is in the range of 7.5~12.1 K and 0.8~2.4 K, respectively. The $T_c$ of H$_2$Br and H$_4$Br is much lower than that of group IVB hydrides, which is mainly attributed to the weaker the EPC.

**Conclusions**

In summary, *ab initio* calculations have been employed to explore the high pressure stability of different stoichiometric H$_n$Br (n=1-7) compounds. Except for H$_6$Br, other stoichiometrics are stable at high pressure. The results show that HBr is stable below 64 GPa, then decomposes into new compound H$_2$Br and Br$_2$ molecular crystal. In addition, H$_2$Br and H$_3$Br were found to become stable above 30 GPa and 8 GPa, respectively. Interestingly, triangular H$_3^+$ species were unexpected found in H$_5$Br compounds at 100 GPa. The $Cmcm$-H$_2$Br and $P6_3/mmc$-H$_4$Br structures are metallic and the predicted $T_c$ of that structures at 240 GPa is 12.1 K and 2.4 K at 240 GPa, respectively. Further experimental studies on the decomposition of HBr and synthesis of these new H-Br compounds at high pressure are in demand.

**Acknowledgements**

This work was supported by the National Basic Research Program of China (No. 2011CB808200), Program for Changjiang Scholars and Innovative Research Team in University (No. IRT1132), National Natural Science Foundation of China (Nos. 51032001, 11204100, 11074090, 10979001, 51025206, 11104102 and 11404134), National Found for Fostering Talents of basic Science (No. J1103202), and China Postdoctoral Science Foundation (2012M511326, 2013T60314, and 2014M561279). Parts of calculations were performed in the High Performance Computing Center (HPCC) of Jilin University.

**Table and Figure Captions**

Figure 1. a-f, Predicted formation enthalpy of $H_nBr$ compounds with respect to decomposition into constituent elemental solids of $Br_2$ and $H_2$ at different pressure. Dashed lines connect data points, and solid lines denote the convex hull. g, Predicted pressure-composition phase diagram of $H_nBr$ compounds.

Figure 2. Selected structures of predicted $H_nBr$ compounds. a, $H_2Br$-*Cmcm*. b, $H_3Br$-*P*-1. c, $H_4Br$-*P*$6_3$/*mmc*. d, $H_5Br$-*Cc*. e, $H_5Br$-*Pmn*$2_1$. f, $H_7Br$-*P*$2_1$/*m*. Large red and small pink spheres represent Br and H atoms, respectively.

Figure 3. The calculated ELF of $H_5Br$. a-c, $H_5Br$-*Cc* at 60, 100 and 140 GPa (plane containing $H_3$, $H_2$ and Br unit), respectively. d, $H_5Br$-*Pmn*$2_1$ at 140 GPa (plane containing $H_3$ and Br unit).

Figure 4. The calculated electronic band structure and DOS projected on Br and H atoms for (a) $H_2Br$-*Cmcm* and (b) $H_4Br$-*P*$6_3$/*mmc* at 240 GPa.

Figure 5. The calculated phonon band structure, projected PHDOS, Eliashberg phonon spectral function $\alpha^2F(\omega)$ and the electron-phonon integral $\lambda(\omega)$ of (a) $H_2Br$-*Cmcm* and and (b) $H_4Br$-*P*$6_3$/*mmc* at 240 GPa.

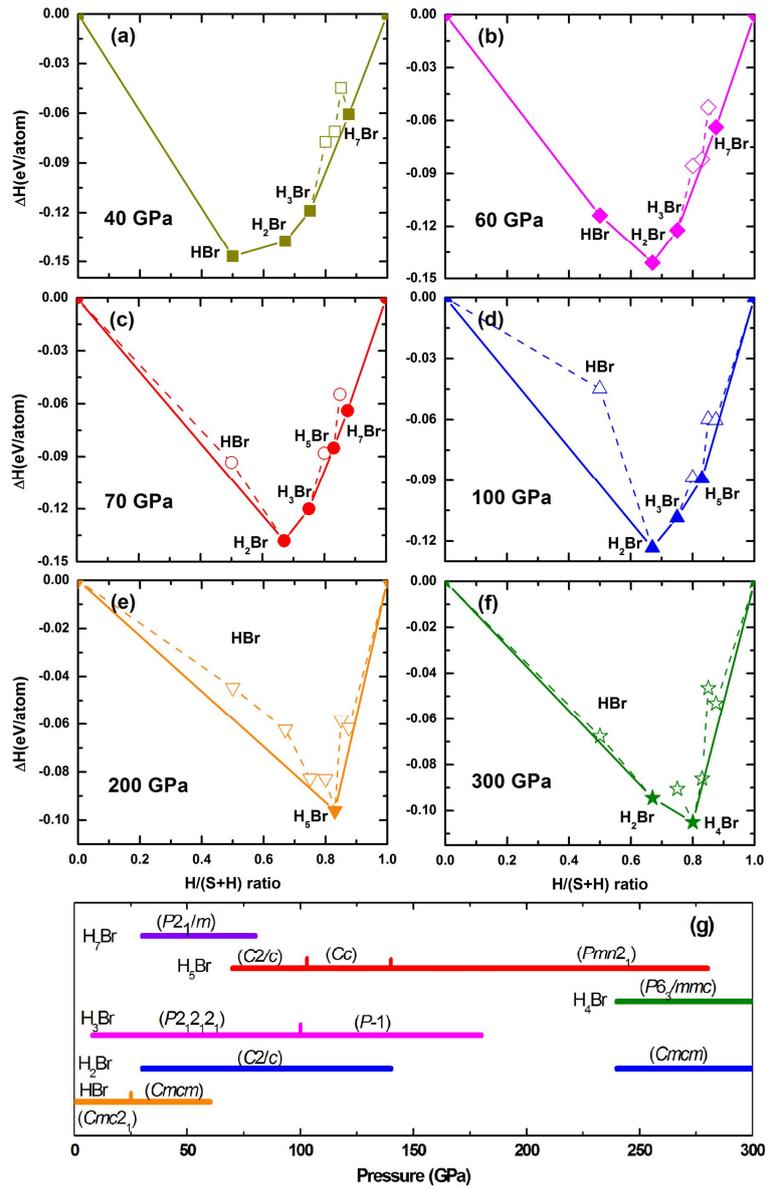

Figure 1

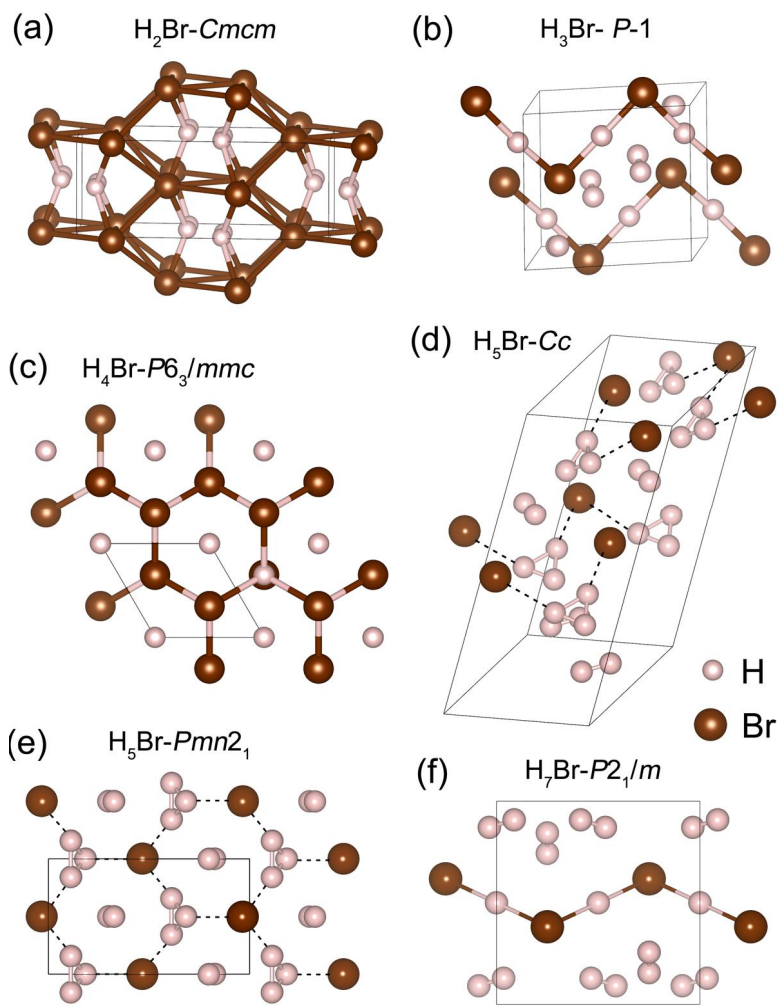

Figure 2

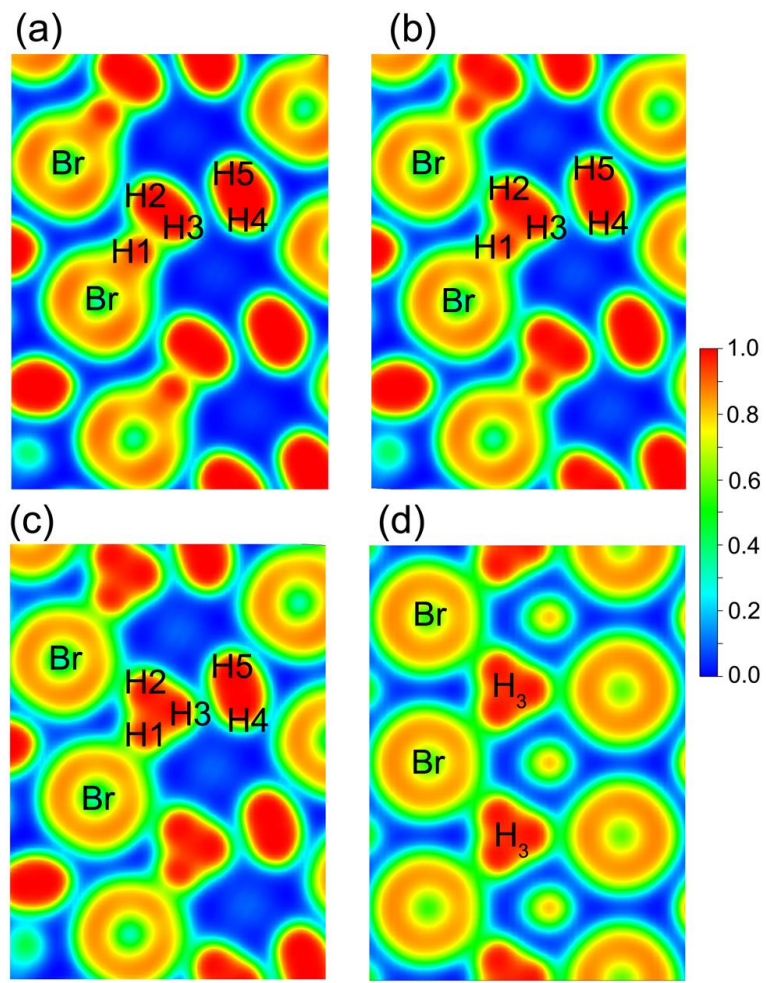

Figure 3

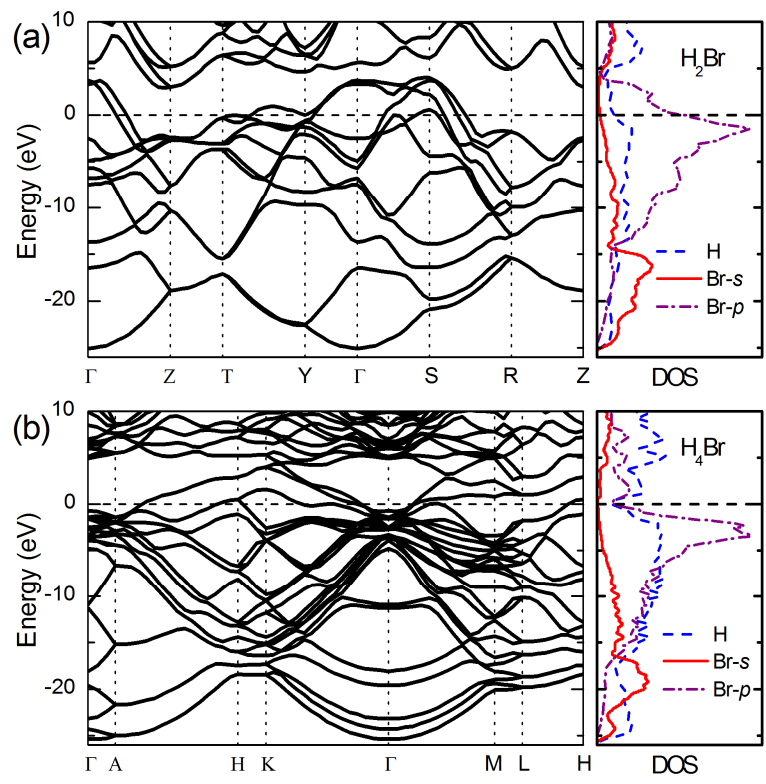

Figure 4

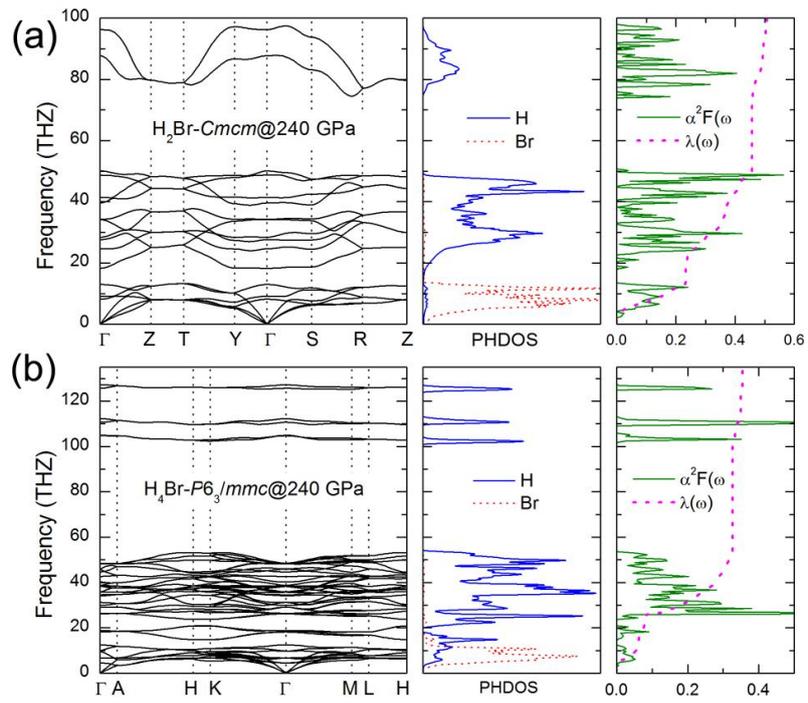

Figure 5

**Supplementary Materials**

# Decomposition of solid hydrogen bromide at high pressure

Defang Duan, Fubo Tian, Xiaoli Huang, Da Li, Hongyu Yu, Yunxian Liu, Yanbin Ma,

Bingbing Liu, Tian Cui*

State Key Laboratory of Superhard Materials, College of Physics, Jilin University,

Changchun 130012, People's Republic of China

*Electronic address: cuitian@jlu.edu.cn

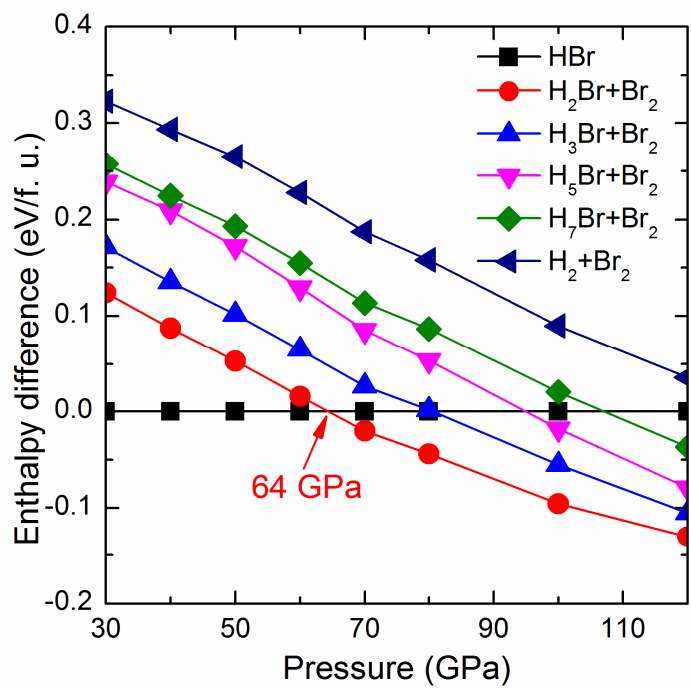

Figure S1 Enthalpies of decomposition of HBr into $H_2Br+Br_2$, $H_3Br+Br_2$, $H_5Br+Br_2$, $H_5Br+Br_2$, and $H_2+Br_2$ as a function of pressure.

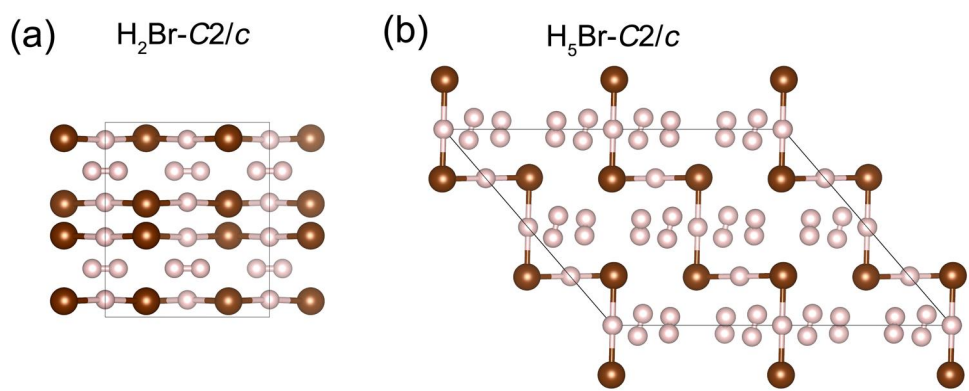

Figure S2 Selected structures of predicted $H_nBr$ compounds. (a), $H_2Br$-*Cmcm* structure. (b), $H_3Br$-*P*-1 structure. (c), $H_5Br$-*C2/c* structure. Large red and small pink spheres represent Cl and H atoms, respectively.

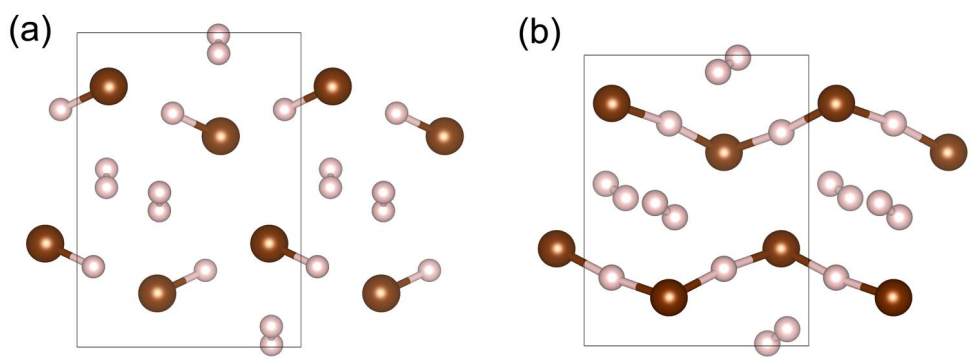

Figure S3 Structures of predicted H$_3$Br-$Pmn2_1$ structure at (a) 10 GPa and (b) 60 GPa. Large red and small pink spheres represent Br and H atoms, respectively.

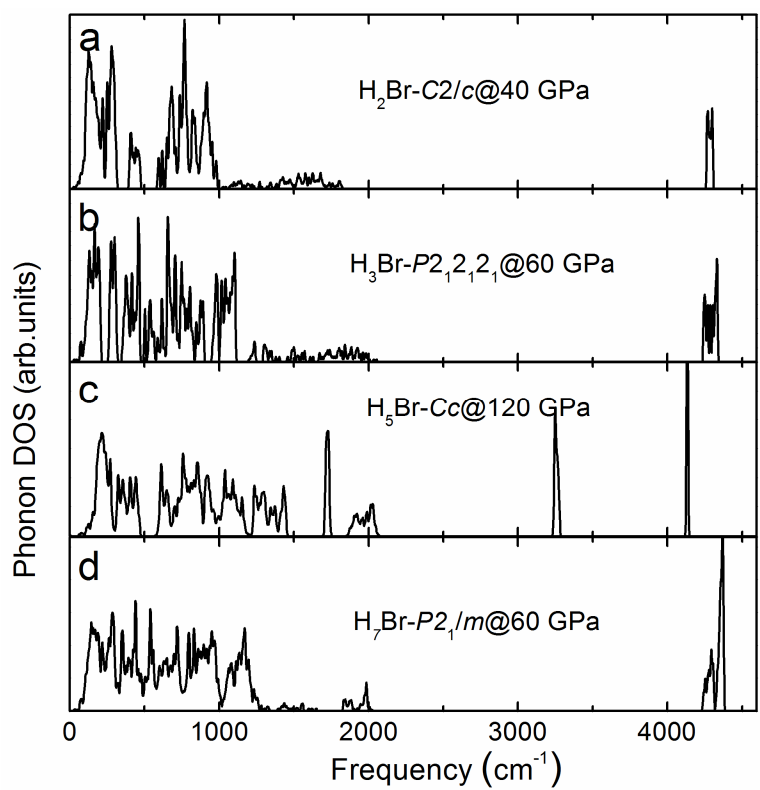

Figure S4 The calculated phonon DOS of selected H$_n$Br compounds.

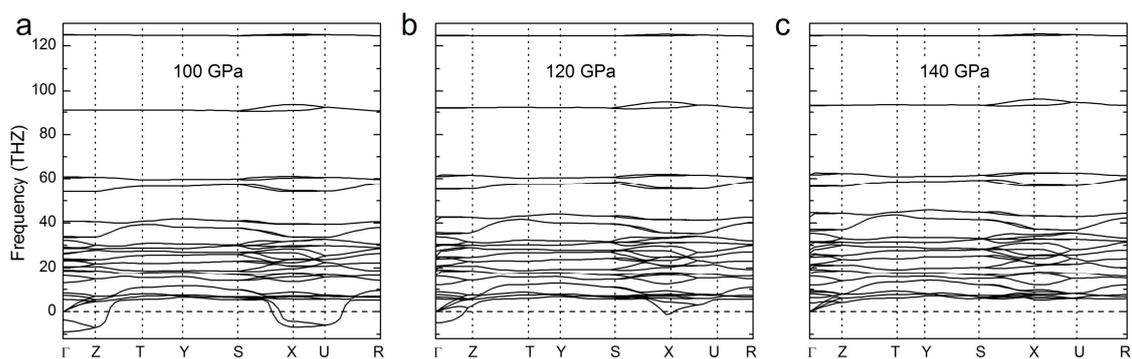

Figure S5 The calculated phonon band structures for $H_5Br$-$Pmn2_1$ at different pressures.

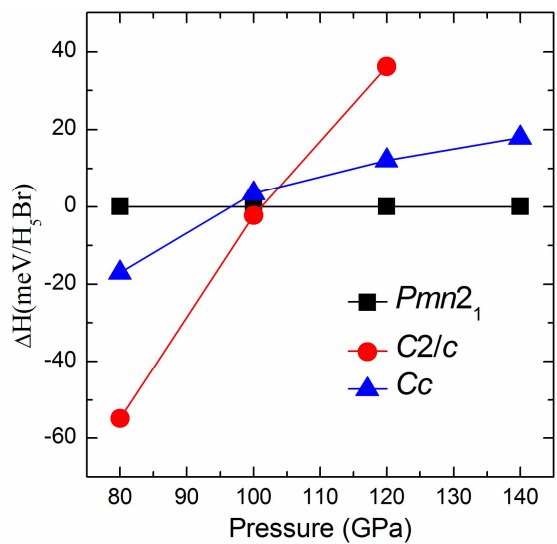

Figure S6 Calculated enthalpy curves (relative to our predicted $Pmn2_1$ phase) for $H_5Br$ as a function of pressure.

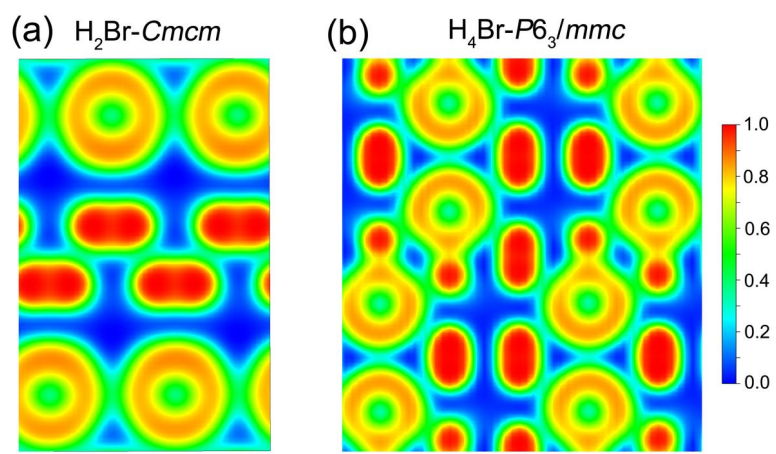

Figure S7 (a) ELF of $H_2Br$-*Cmcm* at (001) plane. (b) ELF of $H_4Br$-$P6_3/mmc$ at (001) plane.